\documentstyle[sprocl]{article}

\def\be{\begin{eqnarray}}
\def\ee{\end{eqnarray}}

\begin{document}

\title{TWO TOPICS IN CHIRAL EFFECTIVE LAGRANGIANS}

\author{Hidenaga YAMAGISHI}

\address{4 Chome 11-16-502, Shimomeguro, Meguro, Tokyo, 153, Japan}

\author{Ismail ZAHED}
\address{Department of Physics, SUNY, Stony Brook, New York 11794, USA.\\
 e-mail: zahed@nuclear.physics.sunysb.edu}

\maketitle\abstracts{
In the absence of nucleons, we use
partial wave unitarity, to show that the chiral expansion parameter must 
be close to $p^2/4\pi f_{\pi}^2$ rather than $p^2/16\pi^2 f_{\pi}^2$ as 
previously suggested, where $p$ is a typical pion momentum and $f_{\pi}$ the 
pion decay constant. When nucleons are included, we apply 
the Tani-Foldy-Wouthuysen (TFW) transformation to the pion-nucleon 
effective Lagrangian to obtain an expansion in powers of $1/m_N$ (inverse
nucleon mass). The results are presented up to order ${\cal O} (1/m_N^3)$, 
corresponding to ${\cal O} (p^4)$ in the momentum. In this case partial
wave-unitarity is also lost in about the same range of momenta.
}

\section{Introduction}

Effective Lagrangians have been an important tool in our understanding
of hadronic processes. If we consider processes without baryons or other 
heavy particles for simplicity, then the loop expansion of the effective 
Lagrangian is equivalent to an expansion in the momentum $p$. Georgi and 
Manohar~\cite{GEORGI1} have suggested that the quantitative expansion parameter 
in this case is $p^2/16\pi^2 f_{\pi}^2 \sim (p/1.2 {\rm GeV})^2$, where the 
factor of $16\pi^2$ arises from a one-loop Feynman diagram.

For $p=m_{\pi}$, this gives 1.4 \% accuracy. However, it is seldom in chiral 
physics to achieve an accuracy of this order. The standard tree level result 
for $\pi\pi\rightarrow \pi\pi$ gives scattering lengths which are off by 40 \% and 25 
\%, depending on the isospin and angular momentum~\cite{EXPERIMENT}. In 
$\gamma\gamma\rightarrow \pi^0\pi^0$ where the Born term is absent, 
the one-loop result is off by 25-30 \% \cite{BIJNENS}.

In this note, we show that the true expansion parameter must be closer to
$p^2/4\pi f_{\pi}^2\sim (p/0.33 {\rm GeV})^2$ through a consideration of 
partial wave unitarity for $\pi\pi$ scattering. For $\pi N$ scattering,
we first show that a pion-nucleon Lagrangian can be organized in terms
of the Tani-Foldy-Wouthuysen transformation, and then observe that partial
wave unitarity in the S31 channel yields a bound that is close to the
$\pi\pi$ one.

\section{Partial Wave Unitarity}

In the chiral limit, the 
invariant amplitude $T^I (s, t)$ with isospin $I$ can be decomposed into 
partial waves as
\be
T^I (s, t) = 32\pi \sum_l (2l+1) \, P_l ({\rm cos} \theta )\,
\eta_l^I\, e^{i\delta_l^I} \, {\rm sin} \delta_l^I
\label{1}
\ee
where $s$ and $t$ are Mandelstam variables, $P_l$ Legendre polynomials, 
$\theta$ the scattering angle in the center of mass frame, $\eta_l^I$
the partial-wave inelasticities, and $\delta_l^I$ the partial-wave phase 
shifts. Projecting out the S-wave gives the bound
\be
|T_0^I| \leq 32\pi
\label{2}
\ee
On the other hand the tree result for massless pions is
\be
T^0 (s, t) = T_0^0 (s) = \frac {8k^2}{f_{\pi}^2}
\label{3}
\ee
where $k$ is the pion momentum in the center of mass frame. It follows 
that the unitarity bound for massless pions is
\be
k^2 \leq 4\pi\, f_{\pi}^2
\label{4}
\ee
and accordingly, the chiral loop expansion parameter must be closer to 
$p^2/4\pi f_{\pi}^2$. 

For massive pions, Eqs.~(\ref{2}) and (\ref{3}) are modified to
\be
|T_0^I (s) | \leq 32\pi\,\,\left(1+\frac {m_{\pi}^2}{k^2}\right)^{\frac 12}
\label{5}
\ee
\be
T^0 (s, t) = T_0^0 (s) = \frac {(8k^2 + 7m_{\pi}^2 )}{f_{\pi}^2}
\label{6}
\ee
Since Eq.~(\ref{5}) is monotonically decreasing and Eq.~(\ref{6}) monotonically 
increasing, we obtain the bound
\be
k^2 \leq 5.2 m_{\pi}^2 \sim (0.32\,\,{\rm GeV})^2
\label{7}
\ee
which is numerically close to Eq.~(\ref{4}).

It follows that the loop expansion of conventional chiral perturbation 
theory~\cite{GASSER} must break down before 300 MeV as far as $\pi\pi$ 
scattering is concerned, in agreement with a similar observation in
\cite{HOLSTEIN}.

\section{Tani-Foldy-Wouthuysen Transformation}

For purely pionic processes such as previously discussed, 
the loop expansion of the effective Lagrangian is 
equivalent to an expansion in the momentum $p$. This is no longer true when 
nucleons are put in Gasser et al.~\cite{GSS}. 
To deal with this situation, heavy baryon 
chiral perturbation theory was proposed~\cite{MANOHAR}, and used as a 
$1/m_N$ expansion~\cite{ALL}. In all of this work, the projected 
fields~\cite{GEORGI}
\be
{\bf N}^{\pm}_v (x) = &&e^{im v\cdot x} \frac 12 (1\pm \rlap/{v})\,\,\psi (x)
\label{I1}
\ee
with $v^2=1$, were employed. 

In the nonrelativistic reduction of the Dirac equation, it is perhaps more 
convenient to work with the TFW transformation~\cite{TFW}, rather than 
Eq.~(\ref{I1}). In this note, we apply the TFW transformation to the pion-nucleon 
effective Lagrangian and obtain terms up to ${\cal O} (p^4)$. 

We take the model relativistic pion-nucleon Lagrangian as 
\footnote{The general case can be found in Steele et al.~\cite{STEELE}, 
for which the present 
arguments also apply.}
\be
{\cal L} =&&+ \frac {f_{\pi}^2}4 {\rm Tr} (\partial_{\mu} U\,\,\partial^{\mu} 
U^{\dagger} ) + \frac 14 f_{\pi}^2m_{\pi}^2 {\rm Tr} (U+U^{\dagger})\nonumber\\
&&+\psi^{\dagger} (i\partial_0 - {\cal  H} ) \psi\nonumber\\
{\cal H} = &&+\vec{\alpha}\cdot (\vec{p} +i\vec{\Gamma}) -i\Gamma_0\nonumber\\
&&-ig_A\sigma^i\Delta_i -ig_A \gamma_5\Delta_0 + \beta m_N\nonumber\\
&&+\frac {m_{\pi}^2}{2\Lambda} (\beta (U+U^{\dagger} -2) +\beta\gamma_5 
(U^{\dagger}-U))
\label{I2}
\ee
where $U=\xi^2$ is the chiral field and
\be
\Gamma_{\mu} =\frac 12 [\xi^{\dagger},\partial_{\mu}\xi]\qquad
\Delta_{\mu} =\frac 12 \xi^{\dagger} (\partial_{\mu} U) \xi^{\dagger}
\label{I3}
\ee
$\Gamma_{\mu}$ and $\Delta_{\mu}$ count as ${\cal O} (p)$ and 
$m_{\pi}^2/\Lambda$ as ${\cal O} (p^2)$.
The Lagrangian is standard except for the term proportional to
$m_{\pi}^2/\Lambda$, which is the pion-nucleon sigma term at tree level.
In QCD the quark mass term generates both the pion mass and the sigma term, so 
the two should go together~\cite{STEELE}.

The idea of TFW is to perform a series of unitary transformations so that the 
upper components of $\psi$ are decoupled from the lower components to a given 
order in $1/m_N$. This makes the evaluation of the fermion determinant
\be
{\rm det}(i\partial_0 -{\cal H}) =
\int [d\psi][d\psi^{\dagger}] {\rm exp}(i\int d^4x \psi^{\dagger} (i\partial_0-
{\cal H} )\psi)
\label{I4}
\ee
straightforward. Since we are interested in terms of ${\cal O} (p^4)$ in 
${\cal H}$, we must work to ${\cal O}(1/m_N^3)$. 

Operators which do not connect the upper and lower components will be called 
even; operators which connect upper components only with lower components will 
be called odd. Algebraically, an even operator ${\cal E}$ obeys ${\cal 
E}\beta=\beta{\cal E}$, and an odd operator ${\cal O}$ obeys ${\cal 
O}\beta=-\beta{\cal O}$. We may write
\be
{\cal H} = &&\beta m_N + {\cal E} + {\cal O} \nonumber\\
{\cal E} = &&-i\Gamma_0 -ig_A\sigma^i\Delta_i + \frac {m_{\pi}^2}{2\Lambda}
\beta (U+U^{\dagger} -2)\nonumber\\
{\cal O} = &&\vec\alpha\cdot(\vec{p}+i\vec\Gamma )-ig_A\gamma_5\Delta_0 +
\frac {m_{\pi}^2}{2\Lambda} \beta\gamma_5 (U^{\dagger}-U)
\label{I5}
\ee
We now apply the unitary transformation $\psi=e^{-iS}\psi'$, where $S$
is taken as ${\cal O}(1/m_N)$. Expanding the exponential to the desired order
\be
&&\psi^{\dagger} (i\partial_0 -{\cal H})\psi = {\psi'}^{\dagger}(i\partial_0 
-{\cal H}' ) \psi'\nonumber\\
&&{\cal H}'={\cal H}+ i [S, {\cal H}] -\frac 12 [S, [S, {\cal H}]]\nonumber\\
&&-\frac i6 [S, [S, [S, {\cal H}]]] +\frac 1{24} 
[S, [S, [S, [S, {\beta m_N}]]]]\nonumber\\
&&-\dot{S}-\frac i2 [S, \dot{S}] + \frac 16 [S,[S, \dot{S}]]
\label{I6}
\ee
where the dot denotes the time derivative. To cancel the odd term to ${\cal O} 
(m_N^0)$, we choose $S=-i\beta {\cal O}/2m_N$, which is consistent with our initial 
assumption.

Substitution into Eq.~(\ref{I6}) gives
\be
{\cal H}' = &&\beta m_N + {\cal E}' + {\cal O}'\nonumber\\
{\cal E}'= &&{\cal E} +\frac 1{2m_N} \beta {\cal O}^2 -\frac 1{8m_N^2}
[{\cal O}, [{\cal O}, {\cal E} ]]\nonumber\\
&&-\frac 1{8m_N^3} \beta {\cal O}^4 -\frac i{8m_N^2} [{\cal O}, \dot{\cal O}]
\nonumber\\
{\cal O}'=&&\frac 1{2m_N} \beta [{\cal O}, {\cal E}]-
\frac 1{3m_N^2} {\cal O}^3 -\frac 1{48m_N^3}
[\beta{\cal O}, [{\cal O}, [{\cal O}, {\cal E}]]]\nonumber\\
&&+\frac i{2m_N} \beta\dot{\cal O} -\frac i{48m_N^3} [\beta{\cal O}, [{\cal O}, 
\dot{\cal O}]]
\label{I7}
\ee
The odd term is now ${\cal O}(1/m_N)$. Applying a second unitary transformation
$\psi'={\rm exp} (-\beta{\cal O}'/2m_N)\psi''$
\be
{\cal H}'' = &&\beta m_N + {\cal E}'' + {\cal O}''\nonumber\\
{\cal E}''= &&{\cal E}' -\frac 1{4m_N^3}\beta ([{\cal O}, {\cal E}] + 
i\dot{\cal O})^2\nonumber\\
{\cal O}''=&& \frac 1{2m_N} \beta [{\cal O}', {\cal E}'] +\frac i{2m_N} \beta 
\dot{\cal O}'
\label{I8}
\ee
The odd term is now ${\cal O} (1/m_N^2)$. Applying a third unitary 
transformation $\psi''={\rm exp} (-\beta{\cal O}''/2m_N)\psi'''$
\be
{\cal H}''' = &&\beta m_N + {\cal E}'' + {\cal O}'''\nonumber\\
{\cal O}'''= &&\frac 1{2m_N} \beta[{\cal O}'', {\cal E}''] + \frac i{2m_N}
\beta\dot{\cal O}''
\label{I9}
\ee
The odd term is now ${\cal O}(1/m_N^3)$. Applying a fourth unitary 
transformation $\psi'''={\rm exp} (-\beta{\cal O}'''/2m_N)\psi''''$
\be
{\cal H}'''' = \beta m_N + {\cal E}'' + {\cal O} (\frac 1{m_N^4})
\label{I10}
\ee
so we have an even Hamiltonian to the desired order. 

We may note that the TFW transformations preserve charge conjugation symmetry, 
so the Hamiltonian Eq.~(\ref{I10}) can be used for the nonrelativistic 
$N\overline{N}$ system, in contrast with previous work.

There is one worry, namely that the functional Jacobian of the transformations. 
However, one can plausibly argue they are 1. In the case of the chiral 
anomaly~\cite{FUJIKAWA}, 
the natural basis for expanding $\psi$ was the eigenfunctions
of the massless Dirac operator, which anticommuted with the generator of axial 
transformations $\gamma_5$. In our case the natural basis for expanding $\psi$ 
should be $\beta$ diagonal, since it anticommutes with the generators 
$\beta {\cal O}$, $\beta {\cal O}'$, $\beta {\cal O}''$, and 
$\beta {\cal O}'''$. However, $\beta$ has no zero modes, so the anomaly should 
vanish. 

Counterterms necessary to absorb loop divergences may be derived in the 
standard manner, by using the BPHZ scheme for instance with on-shell 
subtractions. In this way, the nucleon mass appearing in the expansion at tree
level is the renormalized mass. Incidentally, one observes that the 
relativistic one-loop calculations yield terms of order 
$p^2{\rm ln} m_N/m_N^n$~\cite{GSS,ALL}. How these terms are generated after 
nonrelativistic reduction deserves further investigation. Furthermore,
the ${\cal O} (m_N^0)$ term from Eq.~(\ref{I10}) gives
\be
{\cal H}_0 = -i\Gamma_0 -ig_A\sigma^i\Delta_i +\frac {m_{\pi}^2}{2\Lambda}
\beta (U+U^{\dagger} -2)
\label{I11}
\ee
as the improved version of the static model. Does it account for the $\Delta$ ?

In terms of the present construction, the $\pi N$ scattering amplitude can be 
constructed and partial wave unitarity tested~\cite{STEELE}. Explicit 
calculations using the tree results show that the S31 wave gives a bound on 
the pion momentum to be $k\leq 0.28\,\,{\rm GeV}$. This is close to the bound 
established above using $\pi\pi$ scattering.

\section{Conclusions}

In $\pi\pi$ and $\pi N$ scattering,
if we are to reach 300 MeV and beyond, there are three possible courses of 
action. Use generalized chiral perturbation 
theory~\cite{GENERALIZED}. However, this tends to decrease predictive power.
Try unitarization~\cite{UNITARIZATION}. However, this 
breaks crossing symmetry. The
one we favor is the master formula approach developed recently~\cite{MASTER}. 
In this approach, the $\pi\pi$ scattering amplitude is reduced to a sum of
Green's functions and form factors, some of which are measurable. By making 
educated guesses about the unknown pieces, one may test chiral symmetry even at 
$\rho$ energies. 

In a way, in $\pi\pi$ scattering say,
one should not be worried if lowest order predictions of chiral 
symmetry are off by 20 \% rather than 1.4 \% for $p\sim 140$ MeV. However, one 
should be aware that the predictions can be significantly off already for 
$p\sim 300$ MeV.

\section*{Acknowledgements}
The results in this work are dedicated to Mannque Rho for his sixtieth birthday.
Mannque  has inspired us throughout our careers, and we take this opportunity
to thank him for his friendship and support. This work was supported in part 
by the US DOE grant DE-FG-88ER40388.

\end{document}